\documentclass[preprint,aps, preprintnumbers, nofootinbib]{revtex4}

\newcommand{\gsim}[2]{
\setlength{\unitlength}{12pt}
\begin{picture}(1.4,1.)
\put(.7,-0.3){\makebox(0.0,1.)[t]{$>$}}
\put(.7,-0.3){\makebox(0.0,1.)[b]{$\sim$}}
\end{picture}#2}
\begin{document}
\preprint{IRB-TH-15/05}
\preprint{UB-ECM-PF 05/07}

\title{Hint for Quintessence-like Scalars from Holographic Dark Energy}

\author{B. Guberina\footnote{guberina@thphys.irb.hr}}
\affiliation{\footnotesize Rudjer Bo\v{s}kovi\'{c} Institute,
         P.O.B. 180, 10002 Zagreb, Croatia}

\author{R. Horvat\footnote{horvat@lei3.irb.hr}}
\affiliation{\footnotesize Rudjer Bo\v{s}kovi\'{c} Institute,
         P.O.B. 180, 10002 Zagreb, Croatia}

\author{H. \v{S}tefan\v{c}i\' c\footnote{stefancic@ecm.ub.es}\footnote{On 
leave of absence from the Rudjer
Bo\v{s}kovi\'{c} Institute, Zagreb, Croatia}}
\affiliation{\footnotesize Departament d'Estructura i Constituents de la
Mat\'{e}ria, Universitat de 
Barcelona, Diagonal 647, 08028 Barcelona, Catalonia, Spain}

\begin{abstract}

We use the generalized holographic dark energy model, in which both the
cosmological constant (CC) and  Newton's constant $G_N $ are
scale-dependent, to set constraints on the renormalization-group (RG) evolution
of both quantities phrased within quantum field theory (QFT) in a curved
background. Considering the case in which the energy-momentum tensor of
ordinary matter stays individually conserved, we show from the holographic 
dark energy requirement that the RG laws for the CC and $G_N $ are completely
determined in terms of the lowest part of the particle spectrum of an 
underlying QFT. From
simple arguments one can then infer that the lowest-mass fields should have
a Compton wavelength comparable with the size of the current Hubble horizon.
Hence, although the models with the variable CC (or with both the CC and the $G_N $
varying) 
are known to
lead to successful cosmologies without introducing a new light degree of
freedom, we nonetheless find that holography actually brings us back to the
quintessence proposal. An advantage of having two different components of
the vacuum energy in the cosmological setting is also briefly mentioned.     
\end{abstract}

\newpage

\maketitle

Considering a contribution to the vacuum energy only from states that do
exist in a gravitational holographic theory of gravity \cite{1, 2, 3, 4}, 
provides us with the most elegant solution to the (``old'') cosmological 
constant
(CC) problem \cite{5}. The reason behind this miraculous match between theory
and observation lies in the fact that holography limits the true dynamical
degrees of freedom accessible to a system. 

Another important reason to discuss holography in the context of the CC
problem is that it promotes the vacuum energy density to a scale-dependent
quantity, thus potentially giving a chance to understand also the ``new''
CC problem, that is , the ``coincidence problem'' \cite{6}. Indeed, applying
the entropy bound proposed by Bekenstein {\it et al.} \cite{7} to local
QFTs suggests they must break down in an arbitrary large volume. Additional
relationship between the size of the region $L$ (providing an IR cutoff) and the
UV cutoff was proposed by Cohen {\it et al.} \cite{2}, in order to prevent
formation of black holes within the effective field-theoretical description.
The proposed relationship between the UV and IR cutoff results in an upper
bound on the zero-point energy (ZPE) density $\rho_{\Lambda }$. The largest
$\rho_{\Lambda }$ saturating this inequality can be written as 
\begin{equation}
\rho_{\Lambda }(\mu ) \; \simeq \; \mu^{2} \; G_{N}^{-1}(\mu ) \;,
\end{equation}
where $\mu $ represents the IR cutoff. As
pointed by one of us in \cite{8}, the application of the more stringent
bound of Cohen {\it et al.} \cite{2} to conventional QFTs, in
general  promotes 
not only
$\rho_{\Lambda }$, but also  Newton's constant $G_{N}$ to a dynamical quantity.
Accordingly, we phrase Eq. (1), in which both $G_{N}$ and $\rho_{\Lambda }$
are varying, 
as a generalized holographic dark energy
model. Specifying $L$ as the
size of the present Hubble distance $(L = H_{0}^{-1} \simeq 10^{28}
\;\mbox{\rm
cm})$, one immediately arrives at the observed value for the dark energy
density
today $\rho_{\Lambda} \simeq 10^{-47} \;\mbox{\rm GeV}^4 $.

On the other hand, if $\rho_{\Lambda }$ from (1) is considered as an energy
density of a noninteracting perfect fluid (taking also $G_{N}$ to be a
constant), then 
for some choices of the IR cutoff (the inverse of the size of
the region) one fails to recover the equation of state (EOS) for a dark
energy-dominated universe, as noted in \cite{9}. Specifically, choosing for
$L$ to be the size of the observable universe today, {\it i. e.}, the current
Hubble horizon, one finds that matter and dark energy always scale
identically (for  flat space), $\rho_{\Lambda } \sim \rho_m $, thus
hindering a decelerating era of the universe for redshifts $z \gsim \; 0.5$,
a feature confirmed by the observation of the SNe Ia \cite{10}. Very recently 
it was
found \cite{11} that if the large scale is cut off with the proper event future
horizon, the correct EOS for an
accelerated universe might be obtained. The related issues were discussed 
in \cite{12}. 

In the above examples, an obvious modeling of  dark energy of Eq. (1) is
through self-interacting scalar fields, which still behave as a perfect fluid.
We feel, however, that since Eq. (1) was derived using ZPEs, the most
natural interpretation regarding dark energy in Eq. (1) is through the
variable (or interacting) but ``true'' CC, with the EOS $\omega_{\Lambda}
\equiv p_{\Lambda}/{\rho_{\Lambda}}$ being precisely -1 \cite{9}. To compare
such a model with observation, one should however adapt the framework of the
effective EOS, as defined in \cite{13}. How this works for models involving
the ``true'' CC, see in \cite{14, 15}.   

The transfer of energy between the various components in the universe, in the
framework where also the gravitational constant can be time dependent, is
given  by the generalized equation of continuity \footnote{In Eq. (2) the
quantity $G_{N} T_{total}^{\mu \nu }$ is conserved. In a special case where 
$G_{N}$ is static, the 
total energy-momentum tensor is conserved. The possibility of net creation
of energy in the framework of sourced Friedmann equations was studied in the
transplackanian approach to inflation in \cite{16}, and in the holographic
energy density in \cite{17}.} 
\begin{equation}
\dot{G}_{N}(\rho_{\Lambda } + \rho_m ) + G_N \dot{\rho }_{\Lambda } + 
G_N (\dot{\rho }_{m} + 3H\rho_m )  = 0  \;.
\end{equation}
Eq. (2) is valid for pressureless matter and overdots denote time derivatives. 
Notice also that ${\rho_{\Lambda}}$ in Eq. (2) 
will 
be affected not only by  matter, but also by a time-dependent gravitational
coupling.

In the present paper, we use the holographic restriction (1), supplemented
with the generalized equation of continuity, Eq. (2), to constrain the
parameters of the RG evolution in a conventional field-theoretical model in
curved space. Such a model was based on the observation \cite{18}
that even a ``true'' CC in
such theories cannot be fixed to any definite constant (including zero)
owing to the renormalization-group (RG) running effects. The variation of
the CC
arises solely from  particle field fluctuations, without introducing any
quintessence-like scalar fields. Particle contributions to the
RG  running of the CC which are due to vacuum fluctuations of
massive fields have been properly derived in \cite{19}, with a somewhat
peculiar outcome that more massive fields do play a dominant role in the
running at any scale. When the RG running scale $\mu $ is below the lowest
mass in the theory, we can write the RG laws for  $\rho_{\Lambda }$ and
$G_N $ as \cite{19, 20}
\begin{equation}
\rho_{\Lambda } = \sum^{\infty}_{n=0}C_n \mu^{2n} \;, 
\end{equation}
\begin{equation}
G_{N}^{-1} = \sum^{\infty}_{n=0}D_n \mu^{2n} \;.
\end{equation}
The energy scale $\mu $, associated with the RG running and appearing  
in Eqs. (3) and (4), cannot be set (within QFT and standard cosmology) 
from the first 
principles. We assume that both series
converge well and can be well approximated by retaining just a first few
terms. From the studies of the cosmologies with the running $\rho_{\Lambda
}$ and $G_{N}$ in the formalism of QFT in curved spacetime \cite{19, 20, 18} we
know that generally $C_1 \sim m_{max}^2 $, $C_2 \sim N_b - N_f \sim 1$, $C_3
\sim 1/m_{min}^2 $, etc.; $D_0 = M_{Pl}^2 $, $D_1 \sim 1$, $D_2 \sim 1/m_{min}^2 $,
etc.. Here $m_{max}$ and $m_{min}$ denote the largest and the smallest masses
of massive fields in the theory, respectively, and $N_b$ and $N_f$ stand for
the number of bosonic and fermionic massive degrees of freedom in the
theory, respectively. $C_0 $
represents the ground state of the vacuum (coinciding with the IR limit
of the CC here), which, of course, cannot be
unambiguously  set in the theory.   

We set our context by fixing the matter component in Eq. (2) to evolve in a
canonical way, $\rho_m \sim a^{-3}$, {\it i. e.}, that there is no energy
transfer between this component and both the variable vacuum term and the
time-dependent gravitational field. In this framework, Eq. (2) is reduced to
\begin{equation}
G_{N}'(\mu )(\rho_{\Lambda }(\mu ) + \rho_m ) + G_{N}(\mu )\rho_{\Lambda
}'(\mu ) = 0 \;.
\end{equation}
Here the prime denotes  differentiation with respect to the scale $\mu $.
We  show below that the  scale $\mu $ can be univocally  
fixed \footnote{Here, the scale-fixing is purely 
phenomenological \cite{21} and
is obtained from the equations of continuity, like (2) and (5), and not from
the first-principle considerations of quantum gravity.} only in this
framework, provided RG laws
for both quantities are known. Indeed, after inserting the holographic 
dark energy
requirement (1) into (5), we find 
\begin{equation}
\mu = - \frac{G_{N}'(\mu ) \rho_m }{2} \;,
\end{equation}
which means that there is no more freedom in identification of $\mu $
(that is, the IR cutoff in this case) once $G_{N}(\mu )$ is
known. Some scaling properties of
$\rho_{\Lambda }$ and $G_{N}$ as implied by holography can be easily 
inferred from (6). Namely, from the
requirement of the positivity of the scale $\mu $, $\mu > 0$, it is
seen that $G_{N}'(\mu ) < 0$, which consequently means that $\dot{G}_{N}(t)
> 0$, {\it i. e.}, $G_{N}(t)$ increases as a function of cosmic time. Such
a scale dependence implies that the coupling $G_{N}$ is asymptotically free;
a feature exhibited, for instance, by   higher-derivative quantum gravity
models at the 1-loop level \cite{22}. The asymptotic freedom of $G_{N}$ may
also have an effect on the dynamics of galaxy and their rotation curves
\cite{23, 24}. By similar arguments as above, one can show that
$\dot{\rho}_{\Lambda }(t) < 0$, {\it i. e.}, dark energy decreases as a function
of cosmic time. 
   
In the following we apply the requirement from the generalized holographic
dark energy model (1) to the RG laws as given by Eqs. (3) and (4). This 
relates the coefficients $C$'s and $D$'s in the following way:
\begin{equation}
C_0 = 0; \;\;\;\; C_n \simeq D_{n-1}.
\end{equation}

Before embarking on the  discussion of the announced case, $G_{N}=G_{N}(t)$,
$\rho_{\Lambda }=\rho_{\Lambda }(t)$, $\rho_m \sim a^{-3}$, let us briefly
mention the case with $G_{N}=const$.. In this case, one immediately obtains,
\begin{equation}
C_0 = 0; \;\;\;\; m_{max}^2 \simeq M_{Pl}^2 \;,
\end{equation}
with $\rho_{\Lambda } \sim m_{max}^2 \mu^2 $. On the other hand, the
observational data suggest that $\mu_0 \sim H_0 $, where the subscript `0'
denotes the present-day value. 
It is important to note
that this does not fix the scale at $\mu \sim H $, as one might naively
expect. Indeed, from the continuity equation in the case when $\dot{G}_N  =
0$,
\begin{equation}
\dot{\rho }_{\Lambda } +(\dot{\rho }_{m } + 3H\rho_m )  = 0  \;,
\end{equation}
one  easily sees that the scale $\mu $ cannot be univocally fixed. Eq. (9)
implies a continuous transfer of energy from matter to the CC and {\it vice
versa} (depending on the sign of the interaction term). This means that the
energy density of matter will dilute in a rate whose deviation from the
canonical case $\rho_m \sim a^{-3}$ depends decisively on the choice for $\mu
$. The choice $\mu \sim H $ has been  employed in the model \cite{25}. In 
the light of the assumed merging of QFT with quantum gravity, also note 
that $m_{max} \sim
M_{Pl} $ might represent the effective value of mass owing to multiplicities of
particles having masses just below the Planck scale.

Returning to the most interesting case when both $G_{N}$ and $\rho_{\Lambda
}$ are varying, we insert the expression (4) into the scale-fixing relation
(6), to  arrive at the following expression for the scale $\mu $:
\begin{equation}
\mu^2 \simeq \frac{1}{2} \frac{1-\frac{D_1 }{D_{0}^2 }\rho_m }{\frac{D_2
}{D_1 } - \frac{D_1 }{D_0 }} \;.
\end{equation}                        
Using the estimates for the coefficients $D$'s we finally arrive at 
\begin{equation}
\mu^2 \simeq \frac{1}{2} m_{min}^2   (1 - M_{Pl}^{-4} \rho_m ) \;.
\end{equation} 
Regarding Eq. (11), several comments are in order. The value of the scale
$\mu $ as given by (11) is at least marginally acceptable as far as the
convergence of the expressions (3) and (4) is concerned.\footnote{Note that
Eq. (11) is still an order-of-magnitude estimate. For instance, having $N_b
- N_f \sim 100$ would raise the scale $\mu $ in Eq. (11) by a factor of 
ten. In this
case, one should include more terms from the series (3) and (4) to obtain a
consistent expression for $\mu $.} 
In addition, from
$G_{N}'(\mu ) < 0$ we obtain that $D_1 \simeq C_2 > 0$. Eq. (11) shows an
extremely slow variation of the scale with the scale factor (or cosmic
time). Once the RG scale crosses below the lowest mass in the theory, it
effectively freezes at a value $\sim (1/\sqrt{2}) \; m_{min}$. Confronting 
$\rho_{\Lambda }$ as given by (3) with observation, 
with the scale $\mu $ taken from Eq.
(11), we immediately arrive at $m_{min} \sim H_0 \sim 10^{-33} \;\mbox{\rm
eV}$. We thus find quintessence-like particles in the spectrum. This is the
main result of our paper. It is interesting to notice that what holography
actually does is to expand the particle spectrum from either side to the
extremum; on one side the heaviest possible masses lie near the Planck
scale, on the other side the lowest possible masses are given by the lowest
mass scale in our universe, $H_0 $. Moreover, the ``coincidence problem'' is
easy to understand since  $\rho_{\Lambda }$ today is simply given by the
product of squared masses of the particles lying both on the top and
bottom of the spectrum.

Our results suggest that we may have two different contributions to the
vacuum energy in  cosmological settings. This may help to resolve some of
the cosmological problems, like that with the effective phantom phase of
the universe. Such a superaccelerating phase is indicated by the most recent
observational data, (see, {\it e. g. }, \cite{26}). We refer to a recent model
\cite{15}, comprising both the variable CC and  dark energy modeled as a
scalar field, where a temporary phantom phase can be obtained with a
nonphantom scalar field, having EOS larger than -1. Also, this model was
shown to be free from arguments leading to the Big Rip \cite{27} of the
universe.        

In conclusion, we have shown how merging of a model with a variable CC
based on the RG effects from standard QFT, with the concept of 
holographic dark energy
density, results in remarkable consequences for the particle spectrum of the
former theory. Restricting to the case where the matter energy density has
the usual scaling behavior, we have been able to specify  the RG laws
for both the CC and  Newton's constant in such a manner as to obtain   
the univocal results after conjunction with holography. Although the presence of
quintessence-like scalar fields in the QFT approach to dark energy is
redundant and not required for  consistency with observational data, we
have shown that consistency with holographic predictions calls for their 
appearance in the particle spectrum. Also, in the light of the most recent
cosmological data, we have pointed out to a benefit of having two (or
several) different components of the vacuum energy in the cosmological setup.     
Although we are aware of a ``toy''nature of the holographic energy density,
we still feel that our order-of-magnitude estimates may indicate that this 
interpretation of the dark energy problem,  which favors putting different 
approaches together, ought to be an important ingredient of any realistic  
dark energy model. Forthcoming  astrophysical data will put such a scheme
to test.

{\bf Acknowledgments. } This work was supported by the Ministry of Science,
Education and Sport
of the Republic of Croatia under contract No. 0098002 and 0098011, and
partially supported through the Agreement between the Astrophysical
Sector, S.I.S.S.A., and the Particle Physics and Cosmology Group, RBI. H.
\v{S}. is supported by the Secretar\'{i}a de Estado de 
Universidades e Investigaci\'{o}n of the Ministerio de Educaci\'{o}n y 
Ciencia of Spain under the program "Ayudas para movilidad de Profesores de 
Universidad e Investigadores espa\~{n}oles y extranjeros". The work of H.
\v{S}. is also supported by MCYT FPA 2004-04582-C02-01 and
CIRIT GC 2001 SGR-00065.

\end{document}